\documentclass[numreferences]{kluwer}

\usepackage{klucite}
\usepackage[dvips]{graphicx}
\usepackage{amsfonts}

\newcommand{\sech}{\, {\mathrm {sech}}}
\newcommand{\sign}{\, {\mathrm {sign}} }

\begin{document}

\begin{article}

\begin{opening}

\title{On the existence of conformally coupled scalar field hair 
for black holes in (anti-)de Sitter space}

\author{Elizabeth \surname{Winstanley} \email{E.Winstanley@sheffield.ac.uk}}
\institute{Department of Applied Mathematics, The University of
  Sheffield, Hicks Building, Hounsfield Road, Sheffield. S3 7RH. U.K.}

\runningtitle{Conformally coupled scalar field hair}
\runningauthor{E. Winstanley}

\begin{abstract}
The Einstein-conformally coupled scalar field system is studied in the
presence of a cosmological constant. 
We consider a massless or massive scalar field with no additional
self-interaction, and spherically symmetric black hole geometries.
When the cosmological constant is positive, no scalar hair can exist
and the only solution is the Schwarzschild-de Sitter black hole.
When the cosmological constant is negative, stable scalar field hair
exists provided the mass of the scalar field is not too large. 
\end{abstract}

\end{opening}

\section{INTRODUCTION}
\label{sec:intro}

For thirty years relativists have been concerned with the question of the
uniqueness of black hole solutions to the Einstein equations either in
vacuum or coupled to various types of matter. 
The focus of research during this period has frequently been the proof
of uniqueness theorems for various types of black hole geometry
(static or axisymmetric) and different types of matter (especially
electromagnetic fields).
Over the past ten years or so, a plethora of different black hole
geometries have been found in many models, so that attention has
tended to shift towards finding such solutions (hairy black holes).

The Einstein-scalar field system has long been a favourite during this
period. 
As will be outlined below (see section \ref{sec:status}), 
in asymptotically flat space
there are many theorems which rule out
scalar field hair for various couplings to the geometry and different
types of scalar field potential.
In the presence of a cosmological constant, and with minimal coupling
to the geometry, non-trivial solutions to the field equations have
been found, representing black holes with scalar field hair.
In this article we will consider a non-minimally coupled scalar field
with a non-zero cosmological constant, so that the geometry approaches
(anti-)de Sitter (adS/dS) space at infinity 
rather than being asymptotically flat.

Space-times which are asymptotically (anti-)de Sitter have become the
centre of attention in recent years, for several reasons.
Firstly, observations of high-redshift supernovae suggest that the
universe may possess a small, positive cosmological constant 
\cite{perlmutter}.
Theoretically, there has been much interest in the adS/CFT (conformal
field theory) correspondence (see \cite{aharony} for a review) and its more
recent relative the dS/CFT correspondence 
(see, for example, \cite{strominger}).
Finally, a negative cosmological constant is a key ingredient of
``brane world'' scenarios (see, for example, \cite{randall}).
Black holes with these asymptotic geometries play a particularly
important role in all these developments.

We shall be primarily concerned with a conformally coupled scalar
field since in this case the computations are more straightforward.
This is also the coupling which is most interesting from two points of
view.  
Firstly, as outlined in section \ref{sec:status} below, there exists a
non-trivial solution to the Einstein-conformally coupled scalar field
equations in asymptotically flat space, although the physical
interpretation of this solution has been controversial because the
scalar field diverges at the event horizon.
For other values of the coupling, there are no known non-trivial
solutions to the field equations in asymptotically flat space.
Secondly, conformal coupling is the most natural quantum scalar field 
model to study in quantum field theory in curved space.

The paper is structured as follows.
In section \ref{sec:status} we briefly review the various 
no-scalar hair theorems, for both 
minimally and non-minimally coupled scalar fields.
In section \ref{sec:model} we outline the model under consideration
and derive the boundary conditions on the scalar field.
Next, in section \ref{sec:conformal}, we use a conformal
transformation to map the conformally coupled scalar field system to 
one involving a minimally coupled scalar field.  
We carefully elucidate the conditions under which this transformation
is valid.
We are then in a position to study whether or not there are solutions
to the latter system.
We firstly consider the case in which the cosmological constant is
positive (section \ref{sec:dS}), and show, via a simple proof 
(following Bekenstein \cite{bek72a}) that
does not make use of the conformal
transformation, that no solutions are possible. 
In the situation when the cosmological constant is negative, we find
non-trivial solutions in section \ref{sec:AdS}.
The stability of these solutions is investigated in section
\ref{sec:stab}.
Finally, section \ref{sec:conc} includes our conclusions and
discussion of our results.

\section{STATUS OF THE NO-SCALAR HAIR THEOREMS}
\label{sec:status}

In this section we review the status of the no-scalar hair conjecture,
for both minimally and non-minimally coupled scalar fields.
A more comprehensive review, complete with rigorous theorems, can be
found in \cite{heuslerbook}.
A nice summary of the status of the no-hair conjecture in general 
can be found in \cite{bek96}.

The seminal paper on the subject of no-scalar hair theorems
is \cite{bek72a}, in which Bekenstein rules out non-trivial hair for a
massive scalar field $\phi $ minimally coupled to gravity.
The remarkable thing about \cite{bek72a} is the simplicity of the
argument, which uses only the scalar field equation and the geometric
properties of the black hole space-time at the regular event horizon
and at infinity.
It does not require detailed analysis of possible solutions; nor the
assumption of spherical symmetry (although staticity is assumed, the
result can be extended to stationary black holes \cite{bek72b}); 
nor the Einstein equations.
The theorem can be readily extended to scalar fields 
having a convex potential $V(\phi )$, so that
\begin{displaymath}
\phi \frac {dV}{d\phi } \ge 0,
\end{displaymath}
and although Bekenstein's original theorem is concerned only with
asymptotically flat space-times, it can be readily extended to 
spherically symmetric, asymptotically de Sitter geometries 
\cite{toriidS,cai}.
This argument will be exploited in section \ref{sec:dS} to show the
absence of conformally-coupled scalar field hair for
asymptotically de Sitter black holes.

After this early work of Bekenstein, the subject of scalar field hair
goes fairly quiet in the literature.  
The subject of black hole hair received a new lease of life in the
early 1990's by the discovery of non-trivial black hole solutions
in ${\mathfrak {su}}(2)$ Einstein-Yang-Mills and related theories 
(see \cite{volkov} for a comprehensive review of this topic and an
extensive bibliography).  
In the light of this explosion of interest in hairy black holes,
the old question of scalar field hair was re-addressed, and the
original theorems of Bekenstein extended.
Several authors \cite{heusler95,heusler92,sudarsky} 
proved the no-scalar hair
theorem for a positive semi-definite potential $V(\phi )\ge 0$, 
although now with the additional assumption of spherical symmetry,
but still in asymptotically flat space.
These results rule out non-trivial
scalar fields having the ``double well'' Higgs potential.
The Sudarsky theorem \cite{sudarsky} is particularly simple,
but it does not extend to the situation with a cosmological
constant \cite{toriidS}.
In this paper, we shall not make use of the energy-based arguments of
\cite{heusler95,heusler92}, due in part to the subtleties involved in
making appropriate definitions of energy in non-asymptotically flat
space-times, or where there is non-minimal coupling to the geometry.

Bekenstein himself made an important contribution \cite{bek95}
to this expanding
area by proving a no-scalar hair theorem for a very general model
involving many scalar fields,
again assuming spherical symmetry and minimal coupling to the
geometry, but otherwise making no assumptions about the form of the
interactions between the scalar fields.  
The theorem relies on the assumption that the weak energy condition 
is satisfied by the matter fields, i.e. the stress-energy tensor
$T_{\mu \nu }$ satisfies:
\begin{equation}
T_{\mu \nu } u^{\mu } u^{\nu } \ge 0 
\label{eq:weak}
\end{equation} 
for all time-like vectors $u^{\mu }$, and the analysis makes use of
the conservation equations for the stress tensor and the Einstein equations.
The weak energy condition (\ref{eq:weak}) is satisfied 
by a single scalar field with the usual
Lagrangian and a positive semi-definite potential, but it should be
stressed that the result in \cite{bek95} is considerably more general
than this and makes much weaker assumptions.

More recently, scalar fields in models with a non-zero cosmological
constant have been studied.
The inclusion of a positive cosmological constant allows 
minimally-coupled scalar field
hair to exist for positive semi-definite potentials but not convex 
potentials \cite{toriidS}.
However, this hair is unstable.
If a negative cosmological constant is included, stable scalar field hair
is a possibility \cite{toriiAdS,sudarsky1}, in parallel with the situation
in Einstein-Yang-Mills theory (see \cite{ew1}, where the existence of
stable scalar field hair for asymptotically anti-de Sitter black holes
is conjectured).
Since the focus of the current work is models containing a
cosmological constant, we shall not discuss these results in detail
here, but return to them in later sections.  

So far we have considered only scalar fields minimally coupled to gravity.
Amongst non-minimally coupled scalar field theories, of particular
interest is the case of a conformally coupled scalar field, although
more general couplings can arise in scalar-tensor theories of gravity.
Here, as in the minimally coupled case, seminal contributions to this
topic have been made by Bekenstein himself.  
Bekenstein rediscovered \cite{bek74,bek75} a solution to the
conformally coupled scalar field equations previously found by
Bocharova et al \cite{bbm} but unknown in the West.
In this paper, following \cite{bek96}, we shall refer to this solution
as the BBMB black hole.  
This solution represents an extremal black hole, but its physical
interpretation has been the source of some controversy \cite{zannias}
as the scalar field is divergent on the event horizon, although
Bekenstein \cite{bek75} argued that a particle coupled to the scalar
field would experience nothing pathological at the event horizon.  
Furthermore, this solution is unstable \cite{bronnikov}.
This solution was revisited in the early 1990's by Zannias and
Xanthopoulos \cite{xanth} who proved that it is the unique static,
asymptotically flat, solution of the Einstein-conformal scalar
field system.
Although the BBMB black hole
described above violates the letter of the ``no-hair'' conjecture,
since it is a non-trivial solution of the field equations, no new
extra parameters are introduced, the only additional degree of freedom
provided by the scalar field being a choice of sign.
Therefore this solution has only very limited ``hair'', and certainly
does not violate the ``spirit'' of the conjecture.
Very recently, Martinez et al \cite{martinez} have found the
generalization of the BBMB black hole in the presence of a positive
cosmological constant, with the scalar field having a quartic
potential.
In this case the scalar field is regular on and outside the event
horizon, with a pole inside the event horizon.

In the later 1990's efforts were made to extend the uniqueness theorem
of Xanthopoulos and Zannias to other non-minimally coupled scalar
field theories, not just conformal coupling.
Again, Bekenstein led the way, in a paper with Mayo \cite{mayo}.
There the no-scalar hair theorem is proved when the parameter $\xi $
which couples a neutral scalar field to the curvature (for a precise
definition see equation (\ref{eq:action}) in section \ref{sec:model}
below) is either negative or greater than one half.  
This paper leaves open the question of $0<\xi <1/2$, which includes
the conformally coupled case $\xi =1/6$.
Interestingly, the corresponding proof for a charged, non-minimally
coupled scalar field is valid for all $\xi $.
Related theorems for the neutral case and all values of $\xi $ can be
found in two papers by Saa \cite{saa1,saa2}, which however make
various restrictions on the value of the scalar field, whilst the
paper of Bekenstein and Mayo has stronger results in that the 
restrictions on the value of the scalar field required to prove the
theorems are themselves proved, using energy arguments.  
It is striking that energy arguments can be applied to this system, as
the non-minimal coupling means that the weak energy condition
(\ref{eq:weak}), so crucial in the minimally coupled case, is no
longer valid.
The application of energy arguments to this non-minimally coupled
system has not been without controversy \cite{pena}, although there is
additional numerical evidence \cite{pena} that the theorems of
Bekenstein and Mayo are in fact true.  
We will examine whether or not the energy conditions of \cite{mayo}
are satisfied for the solutions we find in section \ref{sec:AdS}.
 
We are now in a position to test the no-scalar hair conjecture for
non-minimally coupled scalar fields in space-times which are no longer
asymptotically flat.
Our model will contain a (positive or negative) cosmological constant
and so we are interested in black holes in (a)dS space.

\section{THE MODEL AND BOUNDARY CONDITIONS}
\label{sec:model}

In this section we will outline the model to be considered in this
paper, derive the corresponding field equations and use these to
elucidate the boundary conditions on the fields. 

Consider the following action, which describes a self-interacting
scalar field $\phi $ with non-minimal coupling to gravity:
\begin{equation}
S=\int d^{4}x \, {\sqrt {-g}} \left[ 
\frac {1}{2}\left( R -2\Lambda \right) 
-\frac {1}{2} \left( \nabla \phi \right) ^{2}
-\frac {1}{2} \xi R \phi ^{2} -V(\phi ) \right] ,
\label{eq:action} 
\end{equation}
where $R$ is the Ricci scalar, $\Lambda $ the
cosmological constant (which may be either positive or negative), 
$\xi $ is the coupling constant, $V(\phi )$ the scalar field
self-interaction potential and 
$\left( \nabla \phi \right) ^{2} = \nabla _{\mu } \phi  
\nabla ^{\mu } \phi $.
For a minimally coupled scalar field, $\xi =0$ and for conformal
coupling (which is the main focus of this paper), $\xi =1/6$ in four
dimensions.
We are particularly interested in the cases where the potential 
$V$ is either zero (massless scalar field, no self-interaction) 
or $V(\phi )=\frac {1}{2} \mu ^{2} \phi ^{2}$ which 
describes a scalar field of mass $\mu $.
Here and throughout this paper, the metric has signature $(-+++)$,
and we will use units in which the gravitational coupling
constant $\kappa ^{2}=8\pi G$ (with $G$ being Newton's constant)
is set equal to unity and $c=1$.

Variation of the action (\ref{eq:action})
yields the following field equations:
\begin{subequation}
\begin{eqnarray}
\left[ 1- \xi \phi ^{2} \right] G_{\mu \nu }
+g_{\mu \nu } \Lambda 
& = &
\left( 1 - 2\xi \right) \nabla _{\mu } \phi \nabla _{\nu }
\phi 
+\left(2 \xi -\frac {1}{2} \right) g_{\mu \nu } \left(
\nabla \phi \right)^{2} 
\nonumber \\
& & 
-2\xi \phi \nabla _{\mu } \nabla _{\nu } \phi 
+2\xi g_{\mu \nu } \phi \nabla ^{\rho }\nabla _{\rho } \phi
\nonumber \\
& & 
- g_{\mu \nu } V(\phi ) ; 
\label{eq:Einstein}
\\
\nabla _{\mu } \nabla ^{\mu } \phi & = & \xi  R\phi +
\frac {dV}{d \phi } .
\label{eq:phi}
\end{eqnarray}
\end{subequation}
It is useful to take the trace of the Einstein equations 
(\ref{eq:Einstein}) to give the Ricci scalar:
\begin{equation}
R = \frac {\left( 1 - 6 \xi \right) \left( \nabla \phi \right) ^{2}
-6 \xi \phi \frac {dV}{d\phi } + 4V(\phi ) + 4\Lambda }{1 -
\xi \left( 1 - 6 \xi \right) \phi ^{2}} ,
\label{eq:Ricciscalar}  
\end{equation}
where we have made use of the scalar field equation (\ref{eq:phi}) to 
substitute for $\nabla _{\mu } \nabla ^{\mu } \phi $ arising in the
expression for $R$.
This equation enables one to eliminate higher-order derivatives of the
metric from the scalar field equation (\ref{eq:phi}).  
In this paper we shall focus primarily on a conformally coupled scalar
field, in which case $\xi = 1/6$ and (\ref{eq:Ricciscalar})
takes a particularly simple form:
\begin{displaymath}
R= 4\Lambda + 4V(\phi ) - \phi \frac {dV}{d\phi };
\end{displaymath}
and the scalar field equation (\ref{eq:phi}) can be written as:
\begin{equation}
\nabla _{\mu } \nabla ^{\mu } \phi =
\frac {2}{3} \left( \Lambda + V(\phi ) \right) \phi 
+\left( 1 - \frac {1}{6} \phi ^{2} \right) \frac {dV}{d\phi }. 
\label{eq:simplephi}
\end{equation}

We consider a spherically symmetric black hole geometry with line 
element
\begin{eqnarray*}
ds^{2} & = & -\left( 1- \frac {2m(r)}{r} - \frac {\Lambda r^{2}}{3}
\right) \exp (2\delta (r)) \, dt^{2}
\nonumber \\ & & 
+ \left( 1 -\frac {2m(r)}{r} -\frac {\Lambda r^{2}}{3} \right) ^{-1}
dr^{2} 
+ r^{2} \, d\theta ^{2} 
+ r^{2} \sin ^{2} \theta \, d \varphi ^{2},
\end{eqnarray*}  
and we assume that the scalar field $\phi $ depends only on the 
radial co-ordinate $r$.
For later convenience, we define the quantity $N(r)$ as
\begin{displaymath}
N(r)= 1 - \frac {2m(r)}{r} - \frac {\Lambda r^{2}}{3}.
\end{displaymath}
The field equations for
this system are (a prime ${}'$ denotes $d/dr$):
\begin{subequation}
\begin{eqnarray}
\frac {2}{r^{2}} \left( 1- \xi \phi ^{2} \right) m'  & = & 
\xi \Lambda \phi ^{2} + \left( \frac {1}{2} - 2 \xi \right) N \phi '^{2}  
-\xi \phi \phi ' N' - 2 \xi N \phi \phi ''
\nonumber \\ & & 
-\frac {4N}{r} \xi \phi \phi '
+ V(\phi ) ;
\label{eq:Einsym1}
\\
\frac {2}{r} \left( 1 -\xi \phi ^{2} \right) \delta ' & = & 
\left( 1- 2\xi \right) \phi '^{2} - 2 \xi \phi \phi ''
+ 2 \xi \phi \phi ' \delta '  ;
\label{eq:Einsym2}
\\
0 & = & 
N \phi '' + \left( N \delta ' + N' + \frac {2N}{r} \right) 
\phi ' - \xi R \phi - \frac {dV}{d \phi } .
\label{eq:phisym}
\end{eqnarray}
\end{subequation}
It is now possible, using (\ref{eq:Ricciscalar}), to eliminate the
Ricci scalar curvature from the scalar field equation 
(\ref{eq:phisym}) and then also eliminate $\phi ''$ from the
right-hand-side of the Einstein equations 
(\ref{eq:Einsym1},\ref{eq:Einsym2}) above.
Writing the equations in this form would make them more amenable to
numerical integration.

We are interested in black hole solutions possessing a regular,
non-extremal event horizon at $r=r_{h}$, close to which the field
variables have the form
\begin{eqnarray*}
N(r) & = & N'(r_{h}) \left( r - r_{h} \right) + O(r-r_{h})^{2} ;
\nonumber \\
\delta (r) & = & \delta (r_{h}) + O(r-r_{h});
\nonumber \\
\phi (r) & = & \phi (r_{h}) + O(r-r_{h}).
\end{eqnarray*}
There may also be a cosmological event horizon at $r=r_{c}\neq r_{h}$
(depending upon the sign of the cosmological constant), 
with similar expansions of the fields nearby. 

At infinity, 
we will assume that the scalar field $\phi $ approaches a constant,
$\phi _{\infty }$, as $r\rightarrow \infty $, and that the geometry
approaches (a)dS, so that $\delta \rightarrow 0$ at infinity. 
We do not necessarily, at this stage, assume that $\phi $ has an
analytic form near infinity, but rather, as observed for minimally
coupled scalar fields \cite{toriidS,toriiAdS},
\begin{equation}
\phi = \phi _{\infty } + O(r^{-k}) 
\label{eq:phiinfinity1}
\end{equation}
for some $k>0$, whose value we shall ascertain shortly.
This means that
\begin{displaymath}
m(r) \rightarrow {\tilde {\Lambda }} r^{3} + M + O(r^{-1}) ,
\end{displaymath} 
so that
\begin{displaymath}
N(r) \rightarrow 1 - \frac {2M}{r} - \frac {\Lambda _{eff}r^{2}}{3}
+O\left( r^{-2} \right) ,
\end{displaymath}
where we have an ``effective'' cosmological constant
\begin{displaymath}
\Lambda _{eff} = \Lambda + 6{\tilde {\Lambda }}.
\end{displaymath}
We will assume that both $\Lambda $ and $\Lambda _{eff}$ are 
non-zero.
At infinity,
\begin{displaymath}
R \rightarrow  4 \Lambda _{eff} + O(r^{-1}),
\end{displaymath}
and therefore the scalar field equation gives, to leading order,
\begin{displaymath}
4 \xi \Lambda _{eff} \phi _{\infty } =
-\frac {dV}{d\phi } (\phi _{\infty }) .
\end{displaymath}
Therefore, in the particular case in which $V(\phi )\equiv 0$,
(and $\Lambda _{eff} \neq 0$),
it must be the case that
\begin{displaymath}
\phi \rightarrow 0  \qquad \qquad
{\mbox {as $r \rightarrow \infty $.}}
\end{displaymath}
In the other case of particular interest to us in this paper,
$V(\phi )=\frac {1}{2} \mu ^{2} \phi ^{2}$, we have
\begin{displaymath}
4 \xi \Lambda _{eff} \phi _{\infty } =
-\mu ^{2} \phi _{\infty },
\end{displaymath}
which means that either $\phi _{\infty }=0$ or, if 
\begin{equation}
4\xi \Lambda _{eff} = - \mu ^{2},
\label{eq:mfix}
\end{equation}
then $\phi _{\infty }$ is arbitrary.
The condition (\ref{eq:mfix}) is applicable only when 
$\Lambda _{eff}<0$, i.e. when the geometry is asymptotically 
anti-de Sitter.

The first Einstein equation now gives, to leading order,
\begin{displaymath}
6{\tilde {\Lambda }}\left( 1- \xi \phi _{\infty }^{2} \right) 
= \xi \Lambda \phi _{\infty }^{2} + V(\phi _{\infty }).
\end{displaymath}
Hence, in the particular case $\phi _{\infty }=0$,
${\tilde {\Lambda }}=0$ and $\Lambda _{eff}=\Lambda $.
On the other hand, for a massive scalar field such that 
(\ref{eq:mfix}) holds,
\begin{displaymath}
\Lambda _{eff} = \Lambda \left( 1 + \xi \phi _{\infty }^{2} 
\right) ^{-1}.
\end{displaymath}
This equation, combined with (\ref{eq:mfix}), fixes the value of
$\phi _{\infty }$ in this case to be such that
\begin{displaymath}
\phi _{\infty }^{2} =- \frac {1}{\xi } 
\left( 1 + \frac {4\xi \Lambda }{\mu ^{2}}
\right) ,
\end{displaymath}
which can hold only if $\mu ^{2} \le -4\xi \Lambda $.
Therefore, for conformally coupled scalar fields for which 
$\xi = 1/6$ (the primary interest in this paper), we have, 
for $\mu ^{2} \le -2\Lambda /3$, either $\phi _{\infty }=0$ or
\begin{equation}
\phi _{\infty }^{2}=-6 \left( 1 + \frac {2\Lambda }{3\mu ^{2}} \right)  .
\label{eq:phiinfinity}
\end{equation}
In section \ref{sec:AdS} we will rule out solutions 
in which $\phi _{\infty }\neq 0$, 
and therefore we will not consider this possibility further.  

Our final task in this section is to determine the rate at which 
$\phi $ vanishes as we approach infinity, i.e. the power $k$ in 
(\ref{eq:phiinfinity1}).
For the minimally coupled scalar field, it is not necessarily the case
that $k$ is an integer (or even real) \cite{toriiAdS}.
We will assume that $\phi _{\infty }=0$ since this is the case for the
solutions we find in later sections.
Substituting the asymptotic form (\ref{eq:phiinfinity1}) into
the scalar field equation (\ref{eq:phisym}) gives the following
equation for $k$:
\begin{displaymath}
0 = k^{2} - 3k + 12 \xi + 3\frac {\mu ^{2}}{\Lambda },
\end{displaymath}
where we have used $V(\phi )= \frac {1}{2} \mu ^{2} \phi ^{2}$.
This has roots, for the conformally coupled ($\xi = 1/6$)  case:
\begin{equation}
k=\frac {3}{2} \pm {\sqrt {\frac {1}{4}- \frac {3\mu ^{2}}{\Lambda }}}
.
\label{eq:kvalues}
\end{equation}
If the cosmological constant $\Lambda $ is negative, then $k$ is
always real, whereas $k$ can be complex if $\Lambda $ is positive and
$\mu ^{2} > \Lambda /12$.
With $\Lambda <0$, both values of $k$ are positive only if 
$\mu ^{2}<-2\Lambda /3$; otherwise there is a negative value of $k$,
which represents a scalar field $\phi $ which does not converge to 
zero at infinity.
Therefore, when we come to consider solutions in anti-de Sitter space
in section \ref{sec:AdS}, we shall restrict our attention
only to those fields having mass $\mu $ such that
$\mu ^{2} < -2\Lambda /3$.
The leading order behaviour of $\phi $ at infinity is given by the
smaller value of $k$, i.e.
\begin{displaymath}
k=\frac {3}{2} - {\sqrt {\frac {1}{4}- \frac {3\mu ^{2}}{\Lambda }}}
.
\end{displaymath}
If $\Lambda >0$, both values of $k$ (\ref{eq:kvalues}) are positive 
when $k$ is real, and
if $k$ is complex, then it always has positive real part.
In this latter situation complex $k$ represents a scalar field
solution which oscillates about zero as $r\rightarrow \infty $,
with the amplitude of the oscillations decreasing like 
$r^{-\Re e \,  k}$.
Similar behaviour is observed for the minimally coupled scalar field
in anti-de Sitter space \cite{toriiAdS}.

\section{CONFORMAL TRANSFORMATION}
\label{sec:conformal}

The system described by the action (\ref{eq:action}) 
above can be transformed 
under a conformal transformation as follows \cite{maeda}:
\begin{equation}
{\bar {g}}_{\mu \nu } =\Omega g_{\mu \nu }
\label{eq:metricT}
\end{equation}
where
\begin{equation}
\Omega = 1-\xi \phi ^{2} .
\label{eq:Omega}
\end{equation}
This transformation is valid only for those solutions for which 
$\Omega $ does not vanish.  
In asymptotically flat space, there is strong numerical \cite{pena}
and analytic \cite{mayo} evidence 
that solutions in which $\Omega =0$ for some value of $r$ do not
exist.
For the solutions to be exhibited in section \ref{sec:AdS}, 
it will be automatic that $\Omega $ has no zeros.
We leave open the question of whether there are in fact solutions in
which $\Omega =0$ at some point.

Under the transformation (\ref{eq:metricT})
the action becomes that of a minimally
coupled scalar field:
\begin{equation}
S=\int d^{4}x \, {\sqrt {-{\bar {g}}}} \left[
\frac{1}{2} \left( {\bar {R}}-2\Lambda  \right)
-\frac {1}{2} \left( {\bar {\nabla }} \Phi \right) ^{2}
-U(\Phi ) \right]
\label{eq:actionT}
\end{equation}
where a bar denotes quantities calculated using the transformed
metric ${\bar {g}}_{\mu \nu }$ and we have defined a new scalar
field $\Phi $ as \cite{maeda}
\begin{equation}
\Phi =\int d\phi \left[ 
\frac {(1-\xi \phi ^{2}) + 6\xi ^{2} \phi ^{2}}{
(1-\xi \phi ^{2})^{2}} \right] ^{\frac {1}{2}} .
\label{eq:Phidef}
\end{equation}
Note that, although the cosmological constant sets a length scale
for the theory, it is unchanged by this conformal transformation.

When $\xi =1/6$, so the scalar field is conformally coupled, the
equation above simplifies to
\begin{eqnarray}
\Phi & = & \int d\phi \, \frac {1}{(1-\phi ^{2}/6)} 
\nonumber 
\\
 & = & {\sqrt {6}} \tanh ^{-1} \left( \frac {\phi }{{\sqrt {6}}} 
\right) .
\label{eq:Phi}
\end{eqnarray}
For all values of $\xi $, we will choose the constant of
integration so that $\Phi =0$ when $\phi =0$.
For values of $\xi $ not equal to $1/6$, the field $\Phi $
(\ref{eq:Phidef}) can be written in terms of inverse
sinh and tanh functions of $\phi $, which means that it is 
not so readily inverted as in the conformally coupled case.
The transformed potential $U(\Phi )$ takes the form \cite{maeda}
\begin{displaymath}
U(\Phi ) = \frac {V(\phi )+\Lambda \xi \phi ^{2} \left(
2- \xi \phi ^{2} \right) }{ \left( 1- \xi \phi ^{2} \right) ^{2}} .
\end{displaymath}

For the remainder of this paper, we shall focus only on the
conformally coupled case, since the conformal transformation 
is straightforward to implement and invert in this case.
The conformally coupled scalar field is also of most interest
physically, as outlined in the introduction. 

\begin{figure}
\begin{center}
\includegraphics[height=4.5in,angle=270]{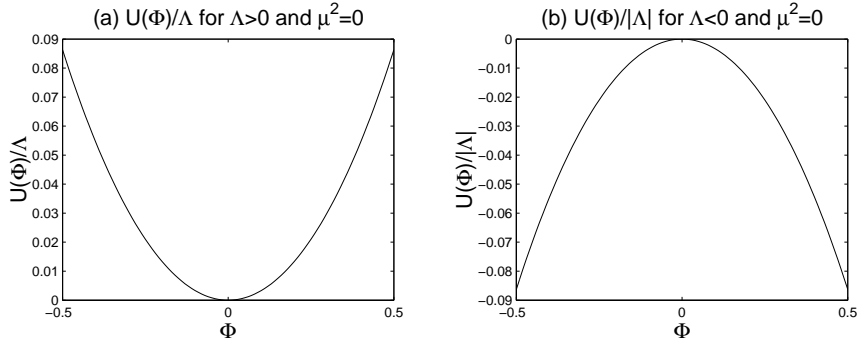}
\caption{(a) Graph of $U(\Phi )/\Lambda $ (vertical axis) against $\Phi $
  (horizontal axis) for positive $\Lambda $ and
  vanishing mass. For a massive conformally coupled scalar field
and $\Lambda >0$, the
  potential has the same overall shape.
(b) Graph of $U(\Phi )/|\Lambda |$ (vertical axis) against 
  $\Phi $ (horizontal axis) for negative $\Lambda $ and
  vanishing mass.}
\label{fig:masslesspot}
\end{center}
\end{figure}

\begin{figure}
\begin{center}
\includegraphics[height=3.5in]{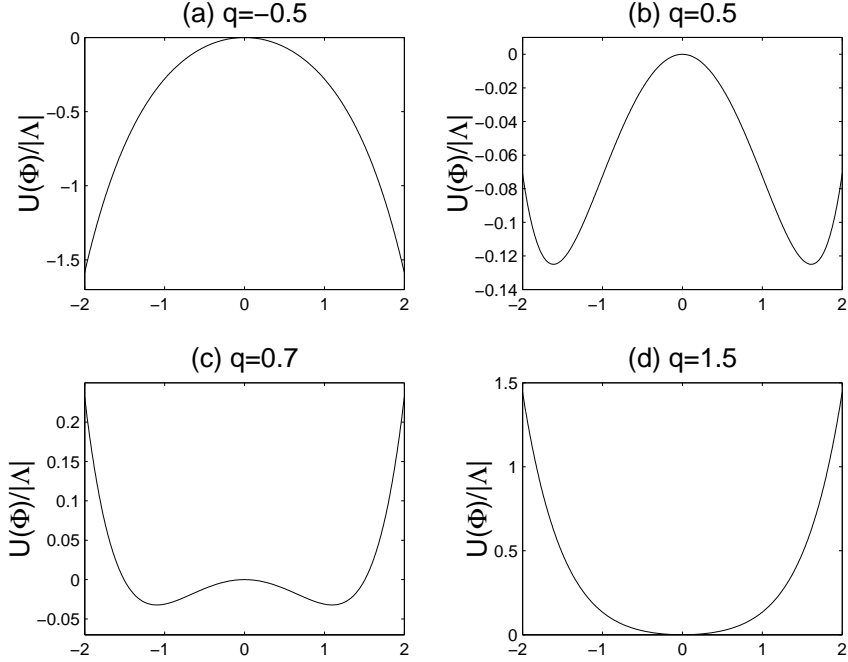}
\caption{Graphs of $U(\Phi )/|\Lambda |$ (vertical axis in each case)
  against $\Phi $ (horizontal axis in each case) for negative $\Lambda $ and
  various values of $q=-1+\frac {3\mu ^{2}}{|\Lambda |}$. For $q\le 0$, the
  potential is negative everywhere as in the massless case (graph
  (a)).  For $0< q < 1$ (graphs (b) and (c)), the potential is 
  negative for small $\Phi $ but becomes positive for sufficiently
  large $\Phi $.  If $q\ge 1$ (graph (d)), then the potential is
  positive for all $\Phi $.}
\label{fig:negLpot}
\end{center}
\end{figure}

Suppose now that the field is conformally coupled, and 
$V(\phi ) \equiv 0$, then
\begin{displaymath}
U(\Phi )  =  \frac {\Lambda \phi ^{2} \left( 
12- \phi ^{2} \right) }{ \left( 6 - \phi ^{2} \right) ^{2} } .
\end{displaymath}
If we use the relation
\begin{equation}
\phi = {\sqrt {6}} \tanh \left( \frac {\Phi }{{\sqrt {6}}} \right) ,
\label{eq:phitrans}
\end{equation}
then
\begin{equation}
U(\Phi ) = \Lambda 
\left[  1+  \cosh ^{2} \left( \frac {\Phi }{{\sqrt {6}}} \right)
\right] 
\sinh ^{2} \left( \frac {\Phi }{{\sqrt {6}}} \right) . 
\label{eq:masslesspot}
\end{equation}
For a massive conformally coupled scalar field, 
$V(\phi )=\frac {1}{2}  \mu ^{2} \phi ^{2}$, the transformed potential is
\begin{eqnarray}
U(\Phi ) & = & \Lambda 
\left[  1+  \cosh ^{2} \left( \frac {\Phi }{{\sqrt {6}}} \right)
\right] 
\sinh ^{2} \left( \frac {\Phi }{{\sqrt {6}}} \right)
\nonumber \\ & &
+ 3 \mu ^{2} \sinh ^{2} \left( \frac {\Phi }{{\sqrt {6}}} \right)
\cosh ^{2} \left( \frac {\Phi }{{\sqrt {6}}} \right) .
\label{eq:massivepot}
\end{eqnarray}
Notice that the sign of the transformed potential $U$ depends on
that of the cosmological constant $\Lambda $.
If $\Lambda >0$, then both the massless (\ref{eq:masslesspot}) and
massive (\ref{eq:massivepot}) potentials are positive,
and have the form sketched in figure \ref{fig:masslesspot} (a).
However, for negative $\Lambda $, the potential corresponding to 
a massless conformally coupled scalar field (\ref{eq:masslesspot})
is always strictly negative and has the form of that for positive
$\Lambda $, but inverted (see figure \ref{fig:masslesspot} (b)).
This is rather unphysical, but it should be borne in mind that
this potential is for the transformed scalar field, which is not 
intended to be physically interpreted.
Rather, the transformation serves to simplify the mathematics, and we
should not be too concerned at this stage that we have an unphysical
transformed potential.
For a massive conformally coupled scalar field, the sign of the potential
(\ref{eq:massivepot}) depends on the mass $\mu $ as well as $\Lambda $.
If $0\le 3\mu ^{2} \le -\Lambda $, then the potential is negative for all
$\Phi $, as for the massless case. 
If $-2\Lambda > 3\mu ^{2} > -\Lambda $, then the potential is positive for
sufficiently large $\Phi $, but negative for small $\Phi $, whilst for
$3\mu ^{2} \ge -2\Lambda $, the potential is positive for all $\Phi $. 
The possibilities are sketched in figure \ref{fig:negLpot}.

The transformed metric ${\bar {g}}_{\mu \nu }$ (\ref{eq:metricT})
will also be spherically symmetric, and we take it to have the form:
\begin{displaymath}
d{\bar {s}}^{2} = -{\bar {N}}({\bar {r}}) 
e^{2{\bar {\delta }}({\bar {r}})} dt ^{2} 
+ {\bar {N}}({\bar {r}}) ^{-1} d{\bar {r}}^{2}
+ {\bar {r}}^{2} \left( d\theta ^{2} + 
\sin \theta ^{2} \, d\varphi ^{2} \right) , 
\end{displaymath}
where we have defined a transformed radial co-ordinate $r$ by
\begin{equation}
{\bar {r}} = \left( 1 - \xi \phi ^{2} \right) ^{\frac {1}{2}} r,
\label{eq:rbar}
\end{equation}
and the transformed metric quantities ${\bar {N}}$ and 
${\bar {\delta }}$ are given by
\begin{eqnarray}
{\bar {N}} & = &
N \left( 1- \xi \phi ^{2} -\xi r \phi \phi' \right) ^{2}
\left( 1- \xi \phi ^{2} \right) ^{-2}  ;
\nonumber \\
e^{\bar {\delta }} & = & e^{\delta } 
\left( 1- \xi \phi ^{2} -\xi r \phi \phi' \right) ^{-1}
\left( 1- \xi \phi ^{2} \right) ^{\frac {3}{2}} .  
\label{eq:metricbar}
\end{eqnarray}
In addition to our earlier requirement that 
$(1-\xi \phi ^{2}) \neq 0$, the transformed radial co-ordinate 
$r$ ceases to be a good co-ordinate if 
\begin{displaymath}
1- \xi \phi ^{2} -\xi r \phi \phi ' =0,
\end{displaymath}
so we are also restricting our attention to those solutions
for which the left-hand-side of the above equation does not vanish.
Using the relationships (\ref{eq:phitrans},\ref{eq:rbar}), 
we see that, in the
conformally coupled case,
\begin{eqnarray}
1-\frac {1}{6} \phi ^{2} & = & 
\sech ^{2} \left( \frac {\Phi }{{\sqrt {6}}} \right) ;
\nonumber \\
1-\frac {1}{6} \phi ^{2} - \frac {1}{6} r \phi \phi ' & = & 
\left[ 1 + \frac {1}{{\sqrt {6}}} {\bar {r}}
\frac {d\Phi }{d{\bar {r}}} \tanh 
\left( \frac {\Phi }{{\sqrt {6}}} \right) 
\right] ^{-1} \sech ^{2}
\left( \frac {\Phi }{{\sqrt {6}}} \right) ,
\label{eq:alttrans}
\end{eqnarray}
so that both conditions are automatically satisfied if 
$\Phi $ is well-defined, in other words if we can find a solution
of the transformed, minimally coupled system, then we automatically
have a solution of the conformally coupled system, provided that
\begin{equation}
{\cal {A}} = 1+
\frac {1}{{\sqrt {6}}} {\bar {r}} 
\frac {d\Phi }{d{\bar {r}}} \tanh \left(
\frac {\Phi }{{\sqrt {6}}} \right) \neq 0 .
\label{eq:bonuscond}
\end{equation}
This condition will be explicitly checked for the solutions we find in
section \ref{sec:AdS}.
The converse, however, is not true, and a solution to the conformally
coupled system does not necessarily lead to a well-defined solution of
the minimally coupled system.  

We may use the equations (\ref{eq:alttrans}) to write 
(\ref{eq:metricbar}) in the alternative form
\begin{eqnarray}
{\bar {N}} & = & 
N \left[ 1 + \frac {1}{{\sqrt {6}}} {\bar {r}} 
\frac {d\Phi }{d{\bar {r}}} \tanh \left(
\frac {\Phi }{{\sqrt {6}}} \right) \right] ^{-2} ;
\nonumber \\
e^{{\bar {\delta }}} & = &
e^{\delta } \left[ 1 + \frac {1}{{\sqrt {6}}} {\bar {r}} 
\frac {d\Phi }{d{\bar {r}}} \tanh \left(
\frac {\Phi }{{\sqrt {6}}} \right) \right] 
\sech \left( \frac {\Phi }{{\sqrt {6}}} \right)  .
\label{eq:metrictrans}
\end{eqnarray}
From (\ref{eq:metricbar},\ref{eq:metrictrans}),
if there is a horizon at for some value of the untransformed
co-ordinate $r$, so that $N(r)=0$, then at the corresponding 
${\bar {r}}$ we will have ${\bar {N}}=0$, so that the transformation
(providing it is regular) does not affect the horizon structure of the
geometry.
At infinity, we have $\phi \rightarrow 0$, and so $\Phi \rightarrow 0$
as well, from our earlier choice of constant of integration in the
definition of $\Phi $.
This means that ${\bar {r}}\sim  r$ as $r\rightarrow \infty $, and
the metric function ${\bar {N}}$ and ${\bar {\delta }}$ behave in
exactly the same way as their untransformed counterparts.

We now define a new quantity ${\bar {m}}({\bar {r}})$ by
\begin{displaymath}
{\bar {N}}({\bar {r}}) = 1 
- \frac {2{\bar {m}}({\bar {r}})}{{\bar {r}}}
-\frac {\Lambda {\bar {r}}^{2}}{3};
\end{displaymath}
in terms of which the field equations derived from the transformed
action (\ref{eq:actionT}) are:
\begin{subequation}
\begin{eqnarray}
\frac {d{\bar {m}}}{d{\bar {r}}} & = & 
\frac {{\bar {r}}^{2}}{4}  {\bar {N}} 
\left(\frac {d\Phi }{d{\bar {r}}}\right) ^{2} 
+ \frac {{\bar {r}}^{2}}{2}  U(\Phi ) ;
\label{eq:minE1}
\\  
\frac {d{\bar {\delta }}}{d{\bar {r}}} & = & 
\frac {{\bar {r}}}{2} \left( \frac {d\Phi }{d{\bar {r}}} \right) ^{2} ;
\label{eq:minE2}
\\
0 & = & 
{\bar {N}} \frac {d^{2}\Phi }{d{\bar {r}}^{2}}  
+ \left( {\bar {N}} \frac {d{\bar {\delta }}}{d{\bar {r}}} 
+ \frac {d{\bar {N}}}{d{\bar {r}}} 
+ \frac {2{\bar {N}}}{{\bar {r}}} \right) 
\frac {d\Phi }{d{\bar {r}}}   
- \frac {dU}{d\Phi } .
\label{eq:mincscal}
\end{eqnarray}
\end{subequation}
From these equations it is clear that the main advantage of
using the conformal transformation is that the minimally coupled
scalar field equations are considerably less complicated than those
for a scalar field with non-minimal coupling, as only first derivatives
of the scalar field now appear on the right-hand-side of the Einstein
equations, and only first derivatives of the metric functions appear
in the scalar field equation.
This means that the minimally coupled scalar field equations are much
easier to solve numerically, although we are restricting our attention
to those solutions for which the conformal transformation is valid.
The other advantage is that the minimally coupled scalar field system
has been extensively studied \cite{toriidS,toriiAdS} and so we can
exploit known results to give us insight into the more complex
non-minimally coupled case.

\section{NON-EXISTENCE OF SOLUTIONS IN DE SITTER SPACE}
\label{sec:dS}

We consider first the case where the cosmological constant $\Lambda $
is positive, so that the geometry is asymptotically de Sitter.

In this situation, the potential in the transformed system 
(\ref{eq:masslesspot},\ref{eq:massivepot}) is
positive semi-definite for both a massless and massive conformally coupled
scalar field (see figure \ref{fig:masslesspot} (a)).
Furthermore, it is convex, since 
\begin{displaymath}
\frac {dU}{d\Phi } = 
\frac {1}{2{\sqrt {6}}} \left( \Lambda + 3 \mu ^{2} \right) 
\sinh \left( \frac {4\Phi }{{\sqrt {6}}} \right) 
+\frac {\Lambda }{{\sqrt {6}}} \sinh 
\left( \frac {2\Phi }{{\sqrt {6}}} \right) ,
\end{displaymath}
which has the same sign as $\Phi $.
The fact that the potential is convex means that it is straightforward 
to prove that there are no non-trivial solutions to the transformed,
minimally coupled system, using Bekenstein's original technique 
\cite{bek72a}.
This argument has been presented in \cite{toriidS}.
However, this above argument is clearly only valid if the conformal
transformation is well-defined everywhere between the event and
cosmological horizons.
In other words, we have assumed that $\phi ^{2}\neq 6$ and 
that $6-\phi ^{2} - r\phi \phi '$ does not vanish anywhere.

In fact, the same technique can be applied directly to the
untransformed system, in the conformally coupled case, to show that
the only solution is $\phi \equiv 0$.
Firstly, we consider the scalar field equation in the simple form
(\ref{eq:simplephi}), for a spherically symmetric geometry in which
\begin{displaymath}
\nabla _{\mu } \nabla ^{\mu } \phi =
r^{-2} e^{-\delta } \left( N r^{2} e^{\delta } \phi ' \right) ' . 
\end{displaymath}
Multiply both sides of the scalar field equation 
by $\phi r^{2} e^{\delta }$ and integrate
from the black hole event horizon $r=r_{h}$ to the cosmological event 
horizon $r=r_{c}$:
\begin{eqnarray}
0 & = & 
\int _{r_{h}}^{r_{c}} dr
\left\{ r^{2} e^{\delta }
\left[ 
\frac {2}{3} \left( \Lambda + V \right) \phi ^{2}
+\left( 1 - \frac {1}{6} \phi ^{2} \right)  
\phi \frac {dV}{d\phi } 
\right] 
\right.
\nonumber \\ & & 
\left.
-
\phi  
\left( Nr^{2} e^{\delta } 
\phi ' \right) ' \right\}
\nonumber \\
& = & 
\int _{r_{h}}^{r_{c}} dr
\, r^{2} e^{\delta } \left[
\frac {2}{3} \left( \Lambda + V \right) \phi ^{2}
+ \left( 1 - \frac {1}{6} \phi ^{2} \right)  
\phi \frac {dV}{d\phi } +
N  \phi  ^{'2} \right]
\nonumber \\
 & & 
-\left[ N r^{2} e^{\delta } 
 \phi \phi ' 
\right] _{r_{h}}^{r_{c}} ,
\label{eq:parts}
\end{eqnarray}
where we have integrated by parts in the second line.
Using the boundary conditions of regular event and cosmological
horizons, the boundary term vanishes.
If the integrand is positive definite, the only possibility is then
that $\phi '\equiv 0$, in which case $\phi \equiv 0$
(using the boundary conditions at infinity), giving 
only the trivial solution.

The integrand in (\ref{eq:parts}) will be positive if the potential
$V$ is positive semi-definite and convex, and if 
$1-\frac {1}{6} \phi ^{2}$ is non-negative everywhere between the 
event and cosmological horizons (which is tantamount to assuming the
validity of the conformal transformation).
However, we are interested in the potentials $V(\phi )=0$  
and $V(\phi )=\frac {1}{2} \mu ^{2} \phi ^{2}$, 
in which case (\ref{eq:parts}) becomes
\begin{equation}
0 = 
\int _{r_{h}}^{r_{c}} dr
\, r^{2} e^{\delta } \left[
\left( \frac {2}{3} \Lambda + \mu ^{2} 
+\frac {1}{6} \mu ^{2} \phi ^{2}  \right) \phi ^{2}
+ N  \phi  ^{'2} \right],
\label{eq:posdefbit}
\end{equation}
so that the integrand is manifestly positive definite and the only
possible solution is the trivial one. 

We note that the fact that we are considering only a conformally
coupled scalar field is crucial in this proof, and for other,
non-minimal couplings, the proof is considerably more complicated.
Furthermore, we are only considering a very restricted class of
potentials (corresponding to massless or massive scalar fields with no
self-interactions) and the proof will not readily extend to more
general potentials (as is the case with the original Bekenstein proof
in asymptotically flat space \cite{bek72a,bek72b}).
The proof given here works equally well for conformal coupling (and
suitably restricted potentials) in asymptotically flat space.
For example, in asymptotically flat space, 
if the potential $V\equiv 0$, then the scalar field 
$\phi $ vanishes like $r^{-1}$ at infinity, whilst if 
$V(\phi ) = \frac {1}{2} \mu ^{2} \phi ^{2} $
then $\phi \sim e^{-\mu r}$ as $r \rightarrow \infty $, so 
that in each case the boundary term at infinity
in (\ref{eq:parts}) vanishes. 
We are therefore left with (\ref{eq:posdefbit}), with $\Lambda =0$,
so that the integrand is manifestly positive definite and the only
possibility is the trivial solution.

Note also that our proof does not rule out the BBMB black hole
\cite{bek74,bek75,bbm}
because the divergence of the scalar field on the (extremal)
event horizon means that the boundary terms in (\ref{eq:parts}) do not
vanish.
For the extension of the BBMB black hole in the presence of a positive
cosmological constant \cite{martinez}, the boundary terms in 
(\ref{eq:parts}) do vanish because the scalar field is now regular on the
event horizon (and the event horizon is no longer extremal).
However, in this case, it is straightforward to check that
the integrand in (\ref{eq:parts}) is no longer
positive definite.

We note, to end this section, that the proof also does not extend to
the case where $\Lambda <0$.
In this situation, in the absence of a cosmological event horizon, 
we would integrate from the black hole event horizon
out to infinity.
From the analysis of section \ref{sec:model}, we are interested only
in values of $\mu $ satisfying $\mu ^{2}<-2\Lambda /3$, in which 
case the behaviour of the scalar field $\phi $ at infinity is such 
that the boundary term in (\ref{eq:parts}) does not vanish,
although it is negative and so makes a positive contribution to the 
right-hand-side of (\ref{eq:parts}).
However, since $\mu ^{2}<-2\Lambda /3$, we can no longer say that the
integrand in the first term of (\ref{eq:parts}), 
i.e. the integrand in (\ref{eq:posdefbit}), is positive definite and 
so the proof fails.
This leaves open the possibility of non-trivial solutions in anti-de
Sitter space, and in the following section we shall show that such
solutions do in fact exist.

\section{EXISTENCE OF SOLUTIONS IN ANTI-DE SITTER SPACE}
\label{sec:AdS}

Having shown that there are no non-trivial solutions to the
conformally coupled scalar field equations in the presence of a
positive cosmological constant, we shall, in this section, 
exhibit numerical solutions when the cosmological constant is
negative, so that the geometry approaches anti-de Sitter space at
infinity.

In this section we shall use the conformal transformation approach
outlined in section \ref{sec:conformal} and are therefore tacitly
assuming that this transformation is valid.  
This method depends on the quantities $1-\phi ^{2}/6$ and 
$1-\phi ^{2}/6-r\phi \phi' /6$ being well-behaved and non-zero
everywhere outside the event horizon.
For the transformation of the fields given in (\ref{eq:Phi}),
both these conditions automatically hold, so that if we can find a
solution in the minimally coupled, transformed system, then we
automatically have a well-behaved solution of the conformally coupled
system. 
The converse is not however true.

We firstly investigate the range of masses for which solutions of this
nature are possible, since the mass alters the form of the 
minimally coupled scalar field potential
(see figure \ref{fig:negLpot}). 
From the analysis of section \ref{sec:model}, 
if $\mu ^{2} \ge -2\Lambda /3$, then $\phi _{\infty }=0$, and the 
scalar field does not have convergent behaviour at infinity.
Therefore, for the remainder of this section, we consider only the 
case $\mu ^{2}\le -2\Lambda /3$.

If $\mu ^{2} \le -2\Lambda /3$, there are two possible behaviours at 
infinity: either $\phi _{\infty }=0$ or $\phi _{\infty } \neq 0$ and
is given by (\ref{eq:phiinfinity}).
In the case that $\phi _{\infty } \neq 0$, the minimally
coupled scalar field $\Phi $ does not tend to zero at infinity
(since in section \ref{sec:conformal} we chose our constant of
integration so that $\Phi =0$ if and only if $\phi =0$).
Working with the minimally coupled scalar field equations means that
we can immediately apply the results of \cite{toriiAdS}.
From \cite{toriiAdS,sudarsky1}, 
it must be the case that $\Phi $ approaches a stationary
point of the potential $U(\Phi )$ at infinity.
For the minimally coupled scalar field potential $U(\Phi )$ given by
(\ref{eq:massivepot}), we have
\begin{eqnarray}
\frac {dU}{d\Phi } & = & 
\frac {1}{2{\sqrt {6}}} \left[ \Lambda + 3 \mu ^{2} \right] 
\sinh \left( \frac {4\Phi }{{\sqrt {6}}} \right) 
+\frac {\Lambda }{{\sqrt {6}}} \sinh 
\left( \frac {2\Phi }{{\sqrt {6}}} \right) ;
\nonumber \\
\frac {d^{2}U}{d\Phi ^{2}} & = & 
\frac {1}{3} \left[ \Lambda + 3 \mu ^{2} \right]
\cosh  \left( \frac {4\Phi }{{\sqrt {6}}} \right)
+\frac {\Lambda }{3} 
\cosh  \left( \frac {2\Phi }{{\sqrt {3}}} \right) .
\label{eq:Uderivs}
\end{eqnarray}
Therefore the potential has stationary points at $\Phi =0$ and 
$\Phi =\pm \Phi _{s}$, where $\Phi _{s}>0$ and
\begin{displaymath}
\cosh \left( \frac {2\Phi _{s}}{{\sqrt {6}}} \right) 
=-\frac {\Lambda }{\Lambda + 3 \mu ^{2}} ;
\end{displaymath}
which is only possible if $-\Lambda /3 \le \mu ^{2}\le -2\Lambda/3 $. 
When $\Phi =0$, we have
\begin{equation}
\frac {d^{2}U}{d\Phi ^{2}} = 
\frac {2\Lambda }{3} + \mu ^{2} ;
\label{eq:d2Uinf}
\end{equation}
and therefore this is a local maximum of the potential (since we are
considering only the masses $\mu ^{2} < -2 \Lambda /3$), 
while the values $\Phi = \pm \Phi _{s}$ correspond to local minima.  
These points can be seen in figure \ref{fig:negLpot}.
In \cite{toriiAdS},
it was found that a local maximum of the potential acted as an
``attractor'' for the scalar field at infinity, and that, while
solutions in which the scalar field tended to a local minimum at
infinity were possible, they necessitated a ``fine tuning'' in the
parameters of the theory and were unstable.
(See also \cite{sudarsky1} for a simple explanation of why the scalar
field must approach a local maximum of the potential at infinity.)
In view of this, we shall from now on only consider solutions in which
$\Phi $ and $\phi $ both tend to zero at infinity.

The minimally coupled field equations of the transformed system 
(\ref{eq:minE1}-\ref{eq:mincscal}), 
are considerably easier to integrate numerically than
the conformally coupled scalar field equations 
(\ref{eq:Einsym1}-\ref{eq:phisym}).
Our approach is to numerically integrate the minimally coupled scalar
field equations outwards from the event horizon for a particular
initial value of the scalar field. 
However, not all initial values of the scalar field at the event horizon,
$\Phi _{h}=\Phi ({\bar {r}}_{h})$, give solutions.
It must be the case that 
$\Phi _{h}\frac {d\Phi }{d{\bar {r}}} ({\bar {r}}_{h}) <0$, which from
the scalar field equation (\ref{eq:mincscal}) implies that
\begin{equation}
\Phi _{h} \frac {dU}{d\Phi }(\Phi _{h}) <0.
\label{eq:initialcond}
\end{equation}
If not, then if $\Phi _{h}>0$, initially the scalar field is
increasing and it will remain in a region for which $dU/d\Phi >0$
(see figure \ref{fig:negLpot}).
If, subsequently, $\Phi $ has a stationary point at 
${\bar {r}}={\bar {r}}_{0}$, then from (\ref{eq:mincscal}),
\begin{displaymath}
{\bar {N}} \frac {d^{2}\Phi }{d{\bar {r}}^{2}} ({\bar {r}}_{0})
= \frac {dU}{d\Phi }(\Phi ({\bar {r}}_{0})) >0,
\end{displaymath}
and so the scalar field can only have a minimum in this region.
Therefore the scalar field is unable to have a maximum, which is
necessary if it is to tend to zero at infinity.
A similar line of reasoning applies if $\Phi _{h}<0$. 
Therefore, using (\ref{eq:Uderivs}) the condition
(\ref{eq:initialcond}) becomes
\begin{equation}
\cosh \left( \frac {2\Phi _{h}}{{\sqrt {6}}} \right) >
-\frac {\Lambda }{\Lambda +3\mu ^{2}} ,
\label{eq:horcond1}
\end{equation}
if $\Lambda + 3 \mu ^{2}<0$, and 
\begin{equation}
\cosh \left (\frac {2\Phi _{h}}{{\sqrt {6}}} \right) <
-\frac {\Lambda }{\Lambda + 3\mu ^{2}} ,
\label{eq:horcond2}
\end{equation}
if $\Lambda + 3 \mu ^{2} >0$.
In the former case, then the right-hand-side of the inequality
in (\ref{eq:horcond1}) is negative and the condition is automatically
satisfied.
However, in the latter case, (\ref{eq:horcond2}) places a restriction
on the value $\Phi _{h}$ of the minimally coupled scalar field at the
event horizon, and the condition can only be satisfied for some 
$\Phi _{h}$ if $3\mu ^{2} < -2\Lambda $ since $\cosh (x)\ge 1$ for all
$x$.
This is not a problem since, as discussed in section \ref{sec:model}
the conformally coupled scalar field has the desired behaviour 
at infinity only if $3\mu ^{2} < -2 \Lambda $.
However, for increasing mass, there is a decreasing interval of 
initial values $\Phi _{h}$ which will give us solutions.

\begin{figure}
\begin{center}
\includegraphics[height=2.5in]{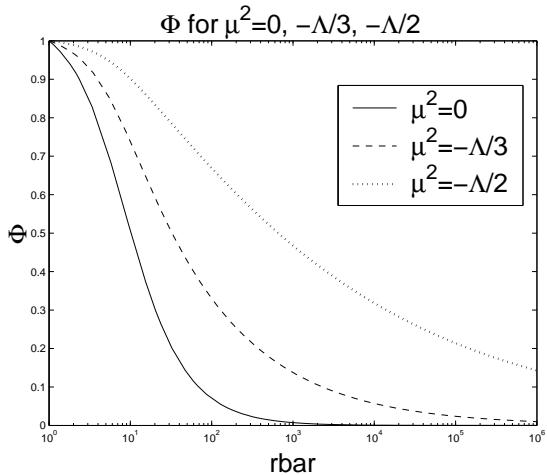}
\caption{Graph of $\Phi $ (vertical axis) against ${\bar {r}} $
(horizontal axis) for $\mu ^{2}=0$,
$-\Lambda /3$ and $-\Lambda /2$, with 
$\Lambda =-0.1$, ${\bar {r}}_{h}=1$ and $\Phi ({\bar {r}}_{h})=1$.
For all values of the mass, $\Phi $ has no zeros and tends to zero at 
infinity.}
\label{fig:Phi}
\end{center}
\end{figure}

With suitable initial conditions, it is then straightforward to 
numerically integrate the minimally coupled scalar field equations
(\ref{eq:minE1}-\ref{eq:mincscal}).
Typical results are shown in figure \ref{fig:Phi}.  
Here we have chosen the particular values $\Lambda = -0.1$, 
${\bar {r}}_{h}=1$, $\Phi ({\bar {r}}_{h})=1$,
and three values of the mass: $\mu =0$,
$\mu ^{2}=-\Lambda /3$  and 
$\mu ^{2} = -\Lambda /2$,
similar behaviour being found for other
values of the parameters.
We plot the scalar field $\Phi $ in figure \ref{fig:Phi}.
We find that the scalar field has no zeros, and simply decays to zero at
infinity, the rate of decay to zero being slower for larger mass.

\begin{figure}
\begin{center}
\includegraphics[height=2.5in]{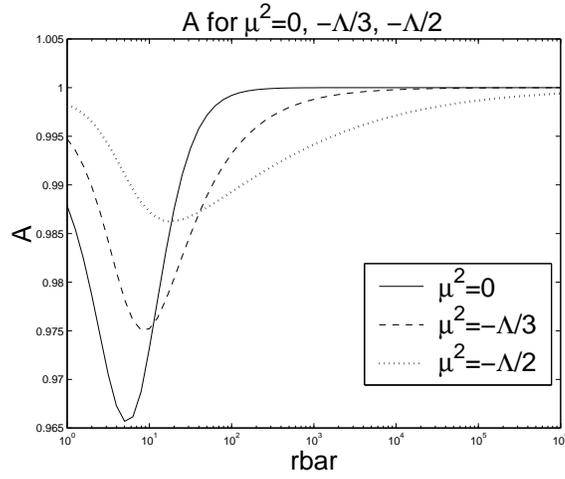}
\caption{Graph of ${\cal {A}}$ (given by (\ref{eq:bonuscond})) 
(vertical axis) against ${\bar {r}} $
(horizontal axis) for $\mu ^{2}=0$, $-\Lambda /3$ and $-\Lambda /2$, with 
the other parameters as in figure \ref{fig:Phi}.
The graph shows clearly that this quantity never
vanishes, so that condition (\ref{eq:bonuscond}) holds and
the conformal transformation remains valid for these
solutions.
Similar results are found for other values of the parameters.}
\label{fig:calA}
\end{center}
\end{figure}

Next, we check that the additional condition 
(\ref{eq:bonuscond}) required for the conformal transformation to be
well-behaved.
As shown in figure \ref{fig:calA} this condition is easily satisfied for 
the solutions we have found.

\begin{figure}
\begin{center}
\includegraphics[height=2.5in]{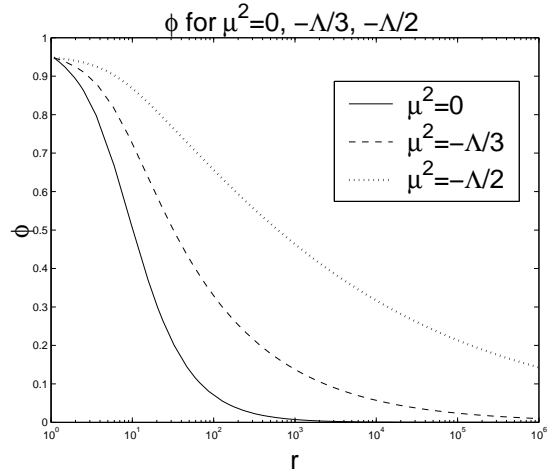}
\caption{Graph of $\phi $ (vertical axis) against $r$ (horizontal axis), 
for the parameter values given
  previously.  The scalar field function is monotonic for
all values of the mass.}
\label{fig:littlephi}
\end{center}
\end{figure}

\begin{figure}
\begin{center}
\includegraphics[height=2in]{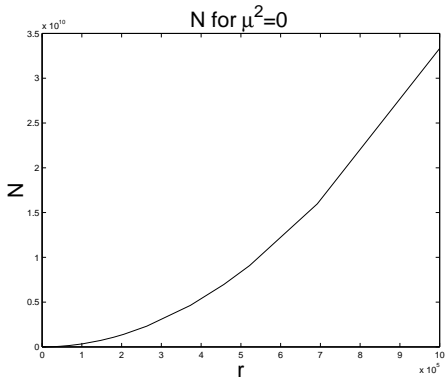}
\caption{Graph of $N$ (vertical axis) against $r$ (horizontal axis), 
for the parameter values given
  previously, just in the specific case $\mu ^{2}=0$.  
The functions for other values of the mass behave similarly.}
\label{fig:N}
\end{center}
\end{figure}

\begin{figure}
\begin{center}
\includegraphics[height=2.5in]{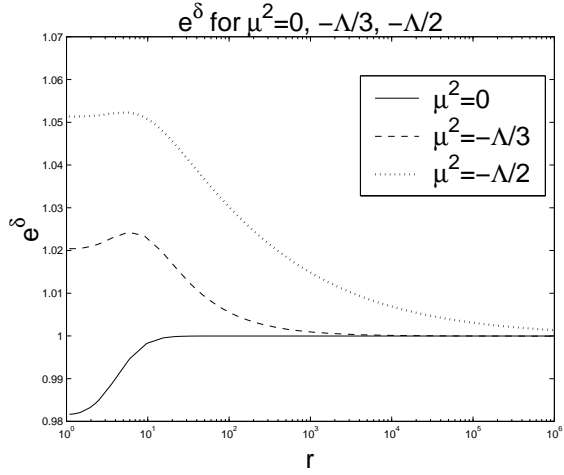}
\caption{Graph of $e^{\delta }$ (vertical axis) against $r$
  (horizontal axis), 
  for the parameter values given
  previously.  In each case $\delta \rightarrow 0$ at infinity, so
  that the geometry is asymptotically anti-de Sitter.}
\label{fig:expdelta}
\end{center}
\end{figure}

We are thus in a position to transform, via
(\ref{eq:phitrans},\ref{eq:rbar},\ref{eq:metricbar}), 
to give the conformally coupled quantities.
In figures \ref{fig:littlephi}-\ref{fig:expdelta} 
we plot the corresponding solutions of the 
conformally coupled scalar field equations for the parameter values
given above.
In common with the minimally coupled scalar field $\Phi $, the
conformally coupled scalar field $\phi $ is monotonic and vanishes
asymptotically at infinity, as can be seen in figure \ref{fig:littlephi}.
This is exactly as is anticipated since $\phi $ has the same sign as 
$\Phi $ from (\ref{eq:phitrans}).
The field decays like $r^{-k}$, where the power $k$ is given by
the smaller of the values (\ref{eq:kvalues}):
\begin{displaymath}
k=\frac {3}{2} - {\sqrt {\frac {1}{4}- \frac {3\mu ^{2}}{\Lambda }}}
.
\end{displaymath}
It is straightforward to check that our numerical solutions have
precisely this power-law fall-off at infinity.

The corresponding metric functions $N(r)$ and $e^{\delta (r)}$ are
plotted in figures \ref{fig:N} and \ref{fig:expdelta} respectively.
Their behaviour is exactly as anticipated, the geometry approaching
anti-de Sitter space as $r\rightarrow \infty $.
Note that, using (\ref{eq:rbar}), the values of the event horizon
radius $r_{h}$ is $1.0845$ for each solution, since we have used the
same value of $\Phi $ at the event horizon.
Similar behaviour is found for other values of the event
horizon radius.
\indent 
The monotonicity of the scalar field $\phi $ is in contrast to some of
the minimally coupled scalar field equations found in \cite{toriiAdS},
where for certain values of the parameters in their double-well
potential, they find solutions in which the scalar field oscillates as
it decays to zero.
Although our potential has, for some values of the parameters, similar
qualitative features as the double-well potential (see figure
\ref{fig:negLpot}) for all values of the parameters of interest in
our case, the scalar field is monotonic.
For hairy black holes in the Einstein-Yang-Mills (EYM) system in adS,
it is found that those solutions in which the single function
describing the gauge field has no zeros are linearly stable
\cite{ew1}.
Stable solutions are also found in the minimally coupled system when
the scalar field has no zeros \cite{toriiAdS}.
One may therefore conjecture that, since in the solutions here the
scalar field has no zeros, these configurations may too be stable.

\begin{figure}
\begin{center}
\includegraphics[height=4in,angle=270]{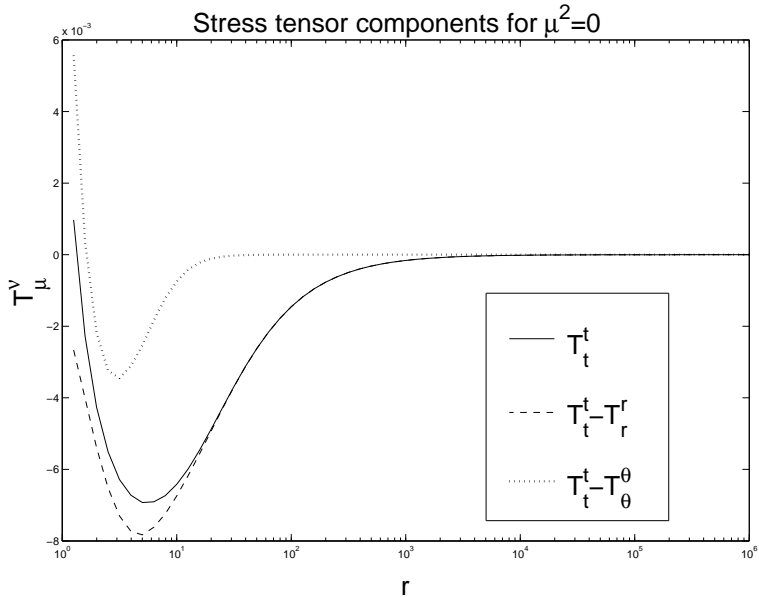}
\caption{Graph of the stress tensor components for the $\mu ^{2}=0$
solution presented above. Note that all three components plotted are
negative except in a region very close to the event horizon.}
\label{fig:stress_0}
\end{center}
\end{figure}

\begin{figure}
\begin{center}
\includegraphics[height=4in,angle=270]{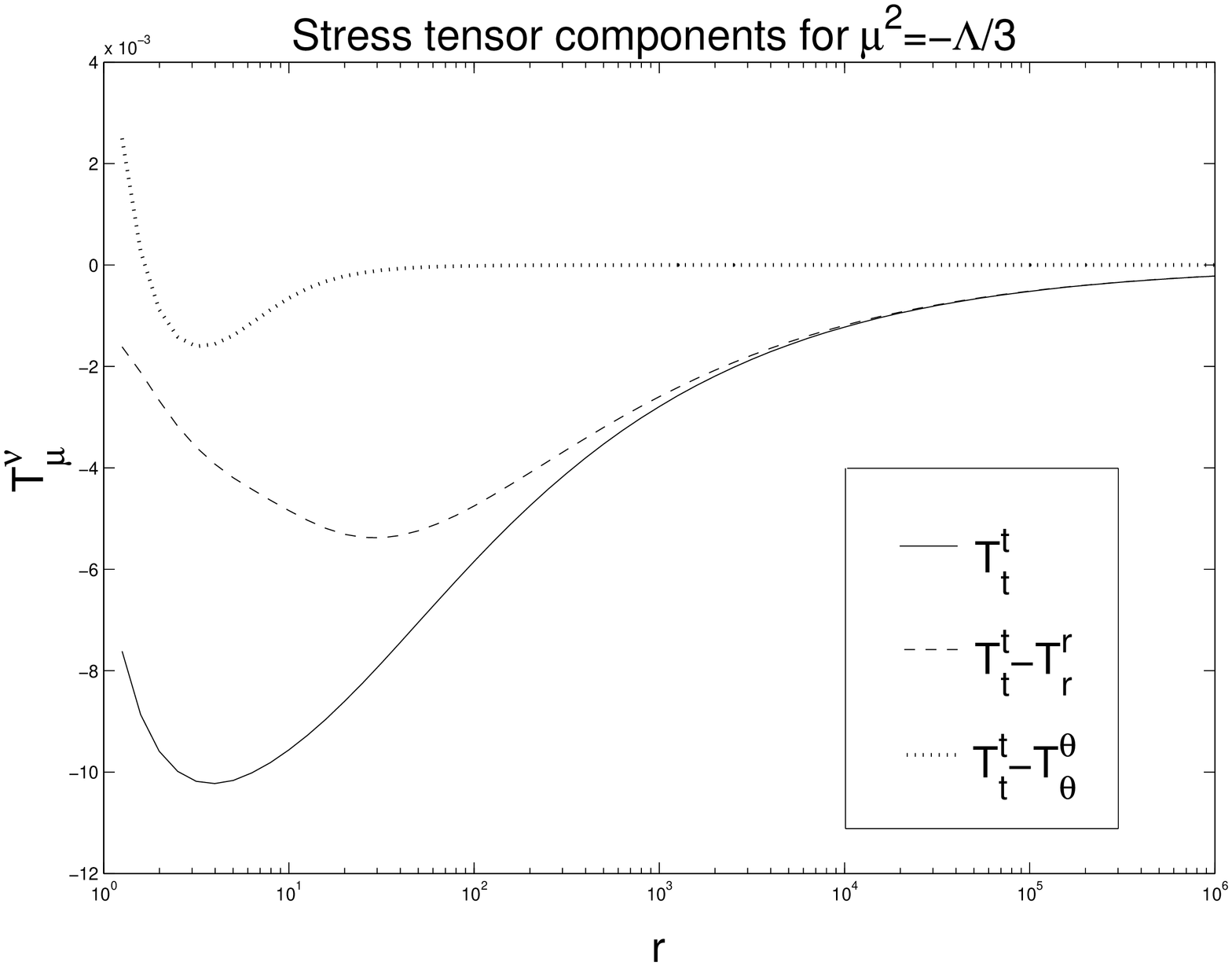}
\caption{As in figure \ref{fig:stress_0}, but with 
$\mu ^{2}=-\Lambda /3$.}
\label{fig:stress_3}
\end{center}
\end{figure}

\begin{figure}
\begin{center}
\includegraphics[height=4in,angle=270]{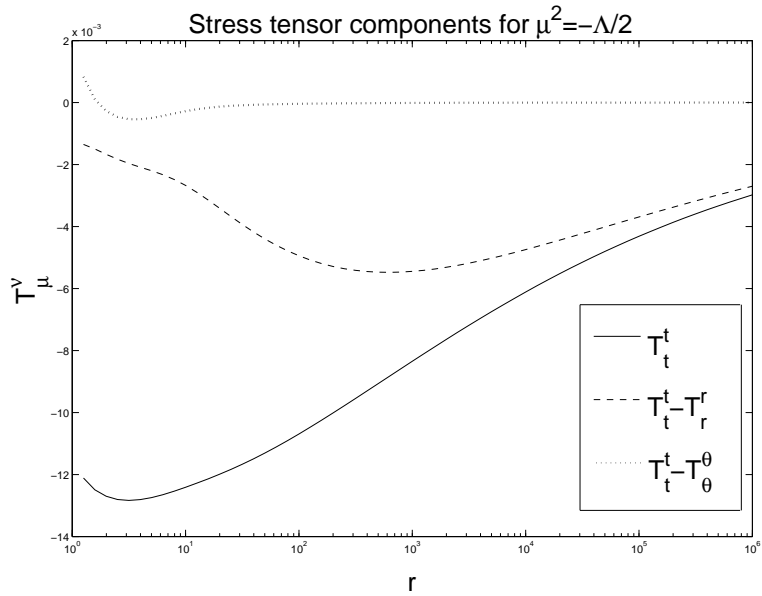}
\caption{As in figure \ref{fig:stress_0}, but with 
$\mu ^{2}=-\Lambda /2$.}
\label{fig:stress_2}
\end{center}
\end{figure}

Before we address this question in the next section, we firstly
consider the stress-energy tensor for the conformally coupled scalar
field.
Its components are given by the following:
\begin{eqnarray}
\left( 1- \frac {1}{6} \phi ^{2} \right)
T_{t}^{t} & = & 
-\frac {1}{6} N \phi ^{'2} 
-\frac {1}{6} e^{-2\delta } \left( Ne^{2\delta } \right) ' \phi \phi '
+\frac {1}{3} \phi \nabla _{\mu }\nabla ^{\mu } \phi 
\nonumber \\ & & 
-V(\phi ) - \frac {1}{6} \Lambda \phi ^{2} ;
\nonumber \\
\left( 1 -\frac {1}{6} \phi ^{2} \right)
\left( T_{t}^{t} - T_{r}^{r} \right) & = & 
-\frac {2}{3} N \phi ^{'2} 
+\frac {1}{3} N \phi \phi ''
-\frac {1}{3} \phi \phi ' \delta ' ;
\nonumber \\
\left( 1 -\frac {1}{6} \phi ^{2} \right)
\left( T_{t}^{t} - T_{\theta }^{\theta } \right) & = & 
\frac {1}{6} \left[ \frac {2N}{r} -
 e^{-2\delta } \left( Ne^{2\delta } \right) ' \right] 
\phi \phi ' .
\label{eq:stress}
\end{eqnarray}
The above formulae are almost identical to those given in \cite{mayo}
for the conformally coupled scalar field in asymptotically flat space.
The cosmological constant simply adds an extra term to $T_{t}^{t}$.
One of the key assumptions of \cite{mayo} is an energy condition that
\begin{equation}
\sign \left( T_{t}^{t} \right) = 
\sign \left( T_{t}^{t} - T_{r}^{r} \right) =
\sign \left( T_{t}^{t} - T_{\theta }^{\theta } \right) .
\label{eq:energycond}
\end{equation}
This is equivalent to the consensus amongst all observers on time-like
paths as to the sign of the energy density.
This is a considerably weaker statement than the weak energy
condition (\ref{eq:weak}),
which states that the three quantities given above are all negative.
It is by no means clear whether this is a reasonable expectation for
our system, given not only the non-minimal coupling of matter to the
geometry, but also the presence of a negative cosmological constant.
We plot, in figures \ref{fig:stress_0}-\ref{fig:stress_2}, 
the three quantities (\ref{eq:stress})
for the solutions presented earlier in this section.
Figures \ref{fig:stress_0}-\ref{fig:stress_2}
show that (\ref{eq:energycond}) holds, 
apart from in a small region close to the event horizon.
In consequence, the weak energy condition is also violated close
to the event horizon.
Violation of the weak energy condition close to the event horizon has
been observed to be a key feature of other models (albeit in
asymptotically flat space) possessing hairy black hole solutions
\cite{kanti,ewnm}.
For all values of the mass, we find that 
$T_{t}^{t}-T_{\theta }^{\theta}$ is positive near the event horizon, 
and that $T_{t}^{t}-T_{r}^{r}$ is negative everywhere outside the
event horizon.
For sufficiently small mass, $T_{t}^{t}$ is positive close to the
event horizon, but for larger values of the mass it is negative
everywhere.
We compare this behaviour with that for the Gauss-Bonnet-dilaton
 black holes in \cite{kanti}, where 
$T_{t}^{t}-T_{r}^{r}$ is always negative,
$T_{t}^{t}-T_{\theta }^{\theta }$ is positive close to the event
horizon, and $T_{t}^{t}$ is positive close to the event horizon.
In both these models there is non-minimal coupling of the scalar field
to the geometry, which seems to be the key factor which results in
the violation of the very weak energy condition 
(\ref{eq:energycond}) near the event horizon. 
This means that, close to the event horizon, there is no consensus
amongst observers as to the sign of the energy density, and some
observers at fixed $r$ (who are highly accelerated) measure a negative
energy density.

The solutions found in this section certainly violate the ``letter''
of the no-hair conjectures, as they represent non-trivial solutions of
the Einstein-conformally coupled scalar field equations.
However, it remains to be seen whether they violate the ``spirit'' of
the conjecture, i.e. whether or not they are stable.
That is the subject of the next section.

\section{STABILITY ANALYSIS}
\label{sec:stab}

We now turn to the question of whether the solutions of the
conformally coupled scalar field equations, exhibited in the previous
section, are stable.

It is algebraically easiest to begin with the transformed, minimally
coupled scalar field system, since we are interested in 
equilibrium solutions
for which the transformation is well-behaved.   
Therefore we consider time-dependent perturbations of the 
equilibrium solutions, keeping the geometry and scalar field
configuration spherically symmetric:
\begin{eqnarray*}
\Phi (t,r) & = & \Phi (r) + \Phi _{1}(t,r); 
\nonumber \\
{\bar {m}} (t,r) & = & {\bar {m}}(r) + {\bar {m}}_{1}(t,r);
\nonumber \\
{\bar {\delta }}(t,r) & = & {\bar {\delta }}(r)
+{\bar {\delta }}_{1}(t,r);
\end{eqnarray*} 
where in each line the first term is the static, equilibrium solution
as discussed in the previous section, and the second term
(with the subscript 1) denotes a small perturbation.
We follow the standard procedure and substitute the above values 
in the time-dependent field equations, 
keeping only terms linear in the perturbations.
It is straightforward to eliminate the metric perturbations, which are
given in terms of the scalar field perturbation $\Phi _{1}$ by
\begin{eqnarray*}
{\bar {m}}_{1} & = & \frac {1}{2}
{\bar {N}} {\bar {r}}^{2}
\frac {d\Phi }{d{\bar {r}}}  \Phi _{1} ;
\nonumber \\
{\bar {\delta }}_{1} & = & 
{\bar {r}} \frac {d\Phi }{d{\bar {r}}} 
\frac {d\Phi _{1}}{d{\bar {r}}} .
\end{eqnarray*}
We consider the usual ``tortoise'' co-ordinate ${\bar {r}}_{*}$ 
given by
\begin{equation}
\frac {d{\bar {r}}_{*}}{d{\bar {r}}} = 
\frac {e^{-{\bar {\delta }}}}{{\bar {N}}} ,
\label{eq:rstardef}
\end{equation}
and end up with a single perturbation equation for $\psi =r\Phi _{1}$,
\begin{equation}
-\frac {\partial ^{2} \psi }{\partial t^{2}} 
= - \frac {\partial ^{2} \psi }{\partial {\bar {r}}_{*}^{2}}
+{\cal {U}} \psi ; 
\label{eq:perteq}
\end{equation}
where the perturbation potential ${\cal {U}}$ is given by
(cf. \cite{toriiAdS}):
\begin{eqnarray}
{\cal {U}} & = & 
\frac {{\bar {N}}e^{2{\bar {\delta }}}}{{\bar {r}}^{2}} 
\left[
1 - {\bar {N}} - \Lambda {\bar {r}}^{2} - U {\bar {r}}^{2} 
+2{\bar {r}}^{3} \frac {dU}{d\Phi } 
\frac {d\Phi }{d{\bar {r}}} 
+ {\bar {r}}^{2} \frac {d^{2}U}{d\Phi ^{2}}
\right.
\nonumber \\
 & & 
\left.  
+ {\bar {r}}^{4} \left( U+\Lambda \right)
\left( \frac {d\Phi }{d{\bar {r}}} \right) ^{2}
-{\bar {r}}^{2} 
\left( \frac {d\Phi }{d{\bar {r}}} \right) ^{2} 
\right]  .
\label{eq:minpertpot}
\end{eqnarray}

It is possible to recast the perturbation equations in terms of
quantities in the untransformed, conformally coupled system.
Using (\ref{eq:metrictrans}), the perturbations of the 
conformally coupled system are given in terms of those in the
minimally coupled system by: 
\begin{eqnarray*}
\phi _{1} & = & \Phi _{1} \sech ^{2}
\left( \frac {\Phi }{{\sqrt {6}}} \right) ;
\nonumber \\
\delta _{1} & = & {\bar {\delta }}_{1} -
\frac {1}{{\sqrt {6}}} \Phi _{1}
\tanh \left( \frac {\Phi }{{\sqrt {6}}} \right) 
\nonumber \\ & & 
-\frac {1}{{\sqrt {6}}} {\bar {r}} 
{\cal {A}}^{-1}
\left[ \frac {d\Phi _{1}}{d{\bar {r}}} 
\tanh \left( \frac {\Phi }{{\sqrt {6}}} \right)
+\frac {1}{{\sqrt {6}}} 
\Phi _{1} \frac {d\Phi }{d{\bar {r}}} 
\sech ^{2} \left( \frac {\Phi }{{\sqrt {6}}} \right) \right]  ;
\nonumber \\
N_{1} & = & 
{\bar {N}}_{1} {\cal {A}}^{2} + 
2{\bar {N}} {\cal {A}} 
\frac {1}{{\sqrt {6}}} {\bar {r}} 
\left[ \frac {d\Phi _{1}}{d{\bar {r}}} 
\tanh \left( \frac {\Phi }{{\sqrt {6}}} \right)
+\frac {1}{{\sqrt {6}}} 
\Phi _{1} \frac {d\Phi }{d{\bar {r}}} 
\sech ^{2} \left( \frac {\Phi }{{\sqrt {6}}} \right) \right] ,
\end{eqnarray*}
where ${\cal {A}}$ is as given in equation (\ref{eq:bonuscond}):
\begin{displaymath}
{\cal {A}} = 
\left[ 1 + \frac {1}{{\sqrt {6}}} {\bar {r}} 
\frac {d\Phi }{d{\bar {r}}} 
\tanh \left( \frac {\Phi }{{\sqrt {6}}} \right) \right]  . 
\end{displaymath}
In the above equations we have considered the perturbations of the
metric function $N$ rather than $m$ as the relationship is simpler to
write down.
In addition, because the radial co-ordinates $r$ and ${\bar {r}}$
are related by (\ref{eq:rbar}), it is no longer the case
that $r$ is unchanged by the perturbations, because we have kept
${\bar {r}}$ fixed during the perturbations of the minimally coupled
system.
This corresponds simply to a choice of gauge in the conformally
coupled system, which, although not the usual choice, does not affect
our conclusions about stability.
Note, however,  
that in transforming back to the conformally coupled system, the
definition of the ``tortoise'' co-ordinate is unchanged, i.e. if
\begin{displaymath}
\frac {dr_{*}}{dr} = \frac {e^{-\delta }}{N},
\end{displaymath}
then $r_{*}={\bar {r}}_{*}$ as a consequence of the transformation
relations (\ref{eq:metricbar},\ref{eq:rstardef}).
Therefore, one could follow the procedure of \cite{bronnikov}, and
rewrite (\ref{eq:perteq}) in terms of quantities in the conformally
coupled system.
However, the numerical values of the potential will be unchanged by
this procedure (although the algebraic form of the potential will
become rather more complicated), and so it is simpler to consider the
potential in the form (\ref{eq:minpertpot}).

\begin{figure}
\begin{center}
\includegraphics[height=3in]{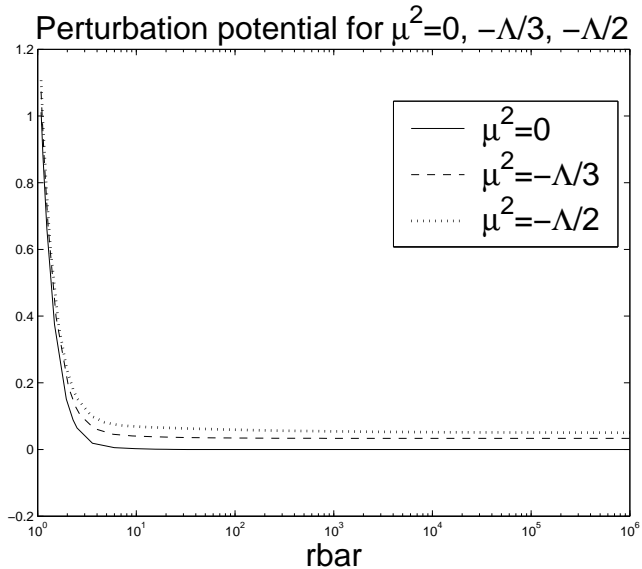}
\caption{Graph of the stability potential 
${\bar {r}}^{2}{\cal {U}}e^{-2{\bar {\delta }}}{\bar {N}}^{-1}$ 
given by (\ref{eq:minpertpot}) for the solutions of section
\ref{sec:AdS}.
Note that in each case the potential ${\cal {U}}$ vanishes at the
event horizon because we have divided by a factor of ${\bar {N}}$ 
and tends to a constant at infinity.}
\label{fig:pot}
\end{center}
\end{figure}

The potential (\ref{eq:minpertpot})
vanishes at the event horizon (${\bar {N}}=0$) 
and its form at infinity can be calculated using (\ref{eq:d2Uinf}) 
and the asymptotic form of the scalar field $\Phi $ to be
\begin{displaymath}
{\cal {U}} \rightarrow  {\bar {N}} \mu ^{2},
\end{displaymath}
which diverges like ${\bar {r}}^{2}$ as 
${\bar {r}} \rightarrow \infty $ due to the factor of ${\bar {N}}$.
For all (non-zero) values of the mass, therefore,
the potential is positive at infinity but we cannot immediately say
whether the solutions are stable or not.
For all the solutions we found,
it was the case that the potential is positive everywhere outside the
event horizon.
Typical examples of such a potential are shown in figure
\ref{fig:pot}, where we have plotted 
${\bar {r}}^{2}{\cal {U}}e^{-2{\bar {\delta }}}{\bar {N}}^{-1}$ 
(i.e. the terms in square brackets in (\ref{eq:minpertpot})) for the 
solutions exhibited in section \ref{sec:AdS}.
The potential ${\cal {U}}$ will vanish at the event horizon because
the functions given in figure \ref{fig:pot} will be multiplied by a
factor of ${\bar {N}}$.
The functions given in figure \ref{fig:pot} are multiplied by a
positive constant at infinity and so the potential ${\cal {U}}$
approaches a positive constant at infinity.
We therefore conclude that the scalar field solutions in this mass
range are linearly stable to spherically symmetric perturbations.
In other words, those solutions in which the scalar field has no zeros
are linearly stable, in common with the EYM system \cite{ew1}.
It may be, of course, that there is some instability due to
non-spherically symmetric perturbations, but we consider this
unlikely.
A very detailed investigation of the EYM system
\cite{olivier1,olivier2} has revealed no instabilities for arbitrary
linear perturbations, provided the cosmological constant is
sufficiently large and negative, and it is reasonable to expect that
the same is true here.

\section{DISCUSSION AND CONCLUSIONS}
\label{sec:conc}

In this paper we have investigated the possibility of giving a black hole
conformally coupled scalar field hair when we remove the restriction
that the geometry be asymptotically flat.
We considered a massless or massive scalar field with potential given
by  $V(\phi )=\frac {1}{2} \mu^{2} \phi ^{2}$, where $\mu $ is the
mass of the field, which may be zero.
For simplicity, we assumed that the scalar field had no other 
self-interactions,
and that the black hole geometry is spherically symmetric.
If the model contains a positive cosmological constant, so that the
geometry approaches de Sitter space asymptotically at infinity, 
we are able to show using a simple method due to Bekenstein
\cite{bek72a} that there are no non-trivial solutions of the field
equations.
With a negative cosmological constant, the geometry is asymptotically
anti-de Sitter and 
we used a conformal transformation to map the conformally coupled
scalar field system to a minimally coupled scalar field system with an
unusual potential.
We numerically integrated the minimally coupled scalar field equations
(which are simpler than those for the conformally coupled scalar
field) and found non-trivial black hole solutions with scalar field
hair, provided that the mass $\mu $ satisfies 
$\mu ^{2} \le -2\Lambda /3$.
These solutions are found to be stable to spherically symmetric linear
perturbations.
If $\mu ^{2} \ge -2\Lambda /3$, then there are no solutions of the
minimally coupled system with the desired boundary conditions.

One might be surprised that we have found a mass bound for conformally
coupled scalar field hair, such that hair exists only for sufficiently
small mass.
However this is in accord with studies of scalar field perturbations
of anti-de Sitter space \cite{breit1,breit2}.
The scalar field equation on this background can
be written as:
\begin{displaymath}
\nabla _{\mu }\nabla ^{\mu } \phi - \alpha \phi =0,
\end{displaymath}
where 
\begin{displaymath}
\alpha = \mu ^{2} + 4 \xi \Lambda 
\end{displaymath} 
is a constant,
for arbitrary coupling to the scalar curvature and arbitrary mass.
It is found that stable fluctuations having positive energy are
possible only for values of $\alpha $ which are bounded above. 
In this situation there is some ambiguity because several different
models (i.e. with different values of mass $\mu ^{2}$ and coupling
constant $\xi $) give the same value of $\alpha $.
However, in our situation in which we have fixed $\xi $, the
conclusion is that $\mu ^{2}$ should be bounded above, as we have found.

Our results should be compared with those for minimally coupled scalar
field hair.
For minimally coupled scalar fields, there are no non-trivial
solutions in asymptotically flat space.
In the presence of a positive cosmological constant, hairy
black holes are possible for non-convex potentials \cite{toriidS} but
these configurations are always unstable.
With a negative cosmological constant, stable scalar field hair is
possible \cite{toriiAdS}.
For conformally coupled scalar fields, the only solution in
asymptotically flat space is the BBMB black hole 
\cite{bek74,bek75,bbm},  which has a
divergent scalar field on the event horizon.
For the simple potentials considered in this paper,
we have shown that there is no conformally coupled scalar field hair
if the cosmological constant is positive,
while stable hair is
possible in the presence of a negative cosmological constant.
Very recently, 
a hairy black hole has been found if the potential is quartic
 \cite{martinez}, generalizing the BBMB black hole (except that 
 now the scalar field is regular on and outside the event horizon).
The stability of this new solution is not yet known, although we think
it is reasonable to conjecture that it is unstable. 
These results are summarised in the table below:
\begin{center}
\begin{tabular}{ccc}
\hline 
 & $\xi =0$ & $\xi =1/6$
\\
\hline 
$\Lambda =0$ & No hair & No regular hair 
\\
$\Lambda >0$ & Unstable hair & (Unstable?) hair 
\\
$\Lambda <0$ & Stable hair & Stable hair 
\\
\hline
\end{tabular}
\end{center}
\noindent
Our main result is that, even within the restricted class of
potentials considered in this paper, 
we have found stable hairy black holes with conformally
coupled scalar field hair if the cosmological constant is negative.

All the solutions we have found are related to solutions of the
minimally coupled scalar field system, via the conformal
transformation (\ref{eq:metricT}).
In asymptotically flat space, there is strong analytic \cite{mayo} and
numerical \cite{pena} evidence that there are no solutions for which
the conformal transformation is not valid.
In asymptotically de Sitter space, the new solution of \cite{martinez}
is such that the conformal transformation breaks down at a point
outside the event horizon.
However, it remains an open question as to whether,
 in asymptotically anti-de Sitter space, all solutions of the
 conformally coupled scalar field system can be obtained from the
 minimally coupled system, or whether there are conformally coupled
 solutions for which the conformal transformation is not valid.

In this paper we have concentrated on a conformally coupled scalar
field because it is the most physically relevant coupling and also the
conformal transformation has its simplest form for this coupling.
The question naturally arises as to whether our results generalise to
other (non-minimal) couplings to the scalar curvature.
In asymptotically flat space, conformal coupling is special in that
only for this coupling is there known to be a non-trivial solution of the 
field equations, namely the Bekenstein black hole, even though the
scalar field diverges at the event horizon.
The theorem of Bekenstein and Mayo \cite{mayo} leaves open the
possibility of solutions for $0<\xi <1/2$, although we do not expect
that solutions will be found for this range of couplings. 
We would conjecture that the results we have presented here are,
qualitatively at least, the same as for more general values of $\xi $,
namely we anticipate that there is stable hair in adS and any hair
that exists in dS is unstable.  
These questions are left for future work. 

\begin{acknowledgements}
We would like to thank Olivier Sarbach for helpful discussions, and 
Ricardo Troncoso and Cristian Martinez
for their helpful comments.  
This work was supported by a grant from the Nuffield Foundation,
reference NUF-NAL 02.
\end{acknowledgements}

\end{article}
\end{document}